\documentclass[twocolumn,showpacs,prl]{revtex4}
\usepackage{amsmath}
\usepackage{graphicx}
\usepackage{dcolumn}
\usepackage{bm}

\begin{document}

\title{ Hopping Transport in Granular Superconductors}

\author{ A.~V.~Lopatin and V.~M.~Vinokur }
\address{ Materials Science Division, Argonne National
Laboratory, Argonne, Illinois 60439, USA }

\date{\today}

\begin{abstract}

We study the conductivity of granular superconductors in the weak
coupling insulating regime. We show that it is governed by the
hopping of either electrons or Cooper pairs depending on the
relation between the superconducting gap and the charging energy
of a single granule. Local superconducting pairing plays an
important role in both cases. In particular, in the case of the
transport via electron hopping the superconducting gap suppresses
the inelastic cotunneling processes. We determine transport
characteristics of an array in different regimes and construct the
transport phase diagram.

\end{abstract}

\maketitle

Recent experiments posed the fundamental questions concerning
mechanisms of conductivity in the insulating phase of
superconductor granular arrays \cite{Gerber}.  The origin of the
Mott variable range hopping - like resistivity temperature
behavior and the nature of strong enhancement of conductivity by
the applied magnetic field in the weak coupling regime are far
from being understood.

At the same time, significant progress has been recently achieved
in understanding the electronic transport in the insulating phase
of {\it metallic} granular arrays. A long standing puzzle of the
low temperature resistivity $\rho(T)$ showing stretched
exponential temperature dependence, $\ln \rho \sim
T^{-1/2}$~\cite{Abeles}, in the insulating phase was explained in
\cite{Zhang05,BLV05,Feigelman05} as the
Mott-Efros-Shklovskii~\cite{ES} (ES) variable range hopping. The
key ingredient of the model of Refs.
\cite{Zhang05,BLV05,Feigelman05} is the effect of the
electrostatic disorder that lifts the Coulomb blockade on a part
of sites of a granular array providing the necessary low energy
electron and hole excitations carrying the current by hopping
through the virtual states of intermediate grains. The specific
hopping mechanism depends on the temperature range
\cite{BLV05,Feigelman05}: at temperatures $T<T_1 \approx 0.1
\sqrt{E_c\delta },$ with $E_c$ and $\delta$ being the charging
energy and the mean energy level spacing in a single grain
respectively, an electron hops via elastic cotunneling mechanism
such that its energy is conserved on each hop. At $T>T_1$ the
inelastic cotunneling mechanism~\cite{Averin} where the travelling
electron creates or absorbs the electron-hole excitations
dominates the transport.  The multiple cotunneling as basic
mechanism of hopping was recently confirmed in experiments on gold
nanoparticle multilayers with controlled structure and size of the
granules~\cite{Jaeger}.

Building on these developments we address in this Letter the low
temperature transport in the insulating phase of granular {\it
superconductors}.  We show that in this case the conductivity is
also of the hopping nature but it is mediated by hopping of either
electrons or Cooper pairs depending on the relation between the
superconducting gap and the charging energy of a single granule.
Electrostatic disorder plays a crucial role in our approach;
models based on {\it regular} superconductor arrays studied
extensively earlier [we refer to the book {\cite{book}} for
review] show activation transport behavior in the insulating
phase.

\begin{figure}[tbp]
\vspace{-0.3cm}
\includegraphics[width=2.6in]{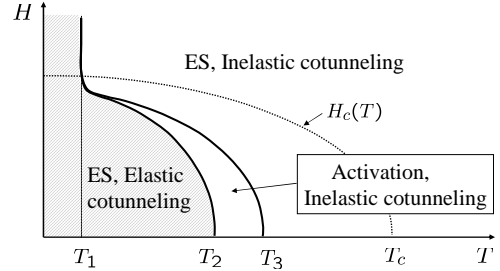}
\caption{ The diagram shows two regions in the EH ( $E_c>\Delta$ )
regime where the hopping conductivity goes via elastic (gray) and
inelastic (white) cotunneling mechanisms in magnetic field $H$ vs.
temperature $T$ plane. The line $H_c(T)$ represents dependence of
the single grain critical field on temperature.
 }
 \vspace{-0.3cm}
  \label{Spectrum1}
\end{figure}
The immediate effect of superconductivity on the properties of
granules is two fold: first, the single particle gap, $\Delta$,
forms within the each grain. Secondly, the so-called {\it parity}
effect \cite{Averin_Nazarov2,Matveev} appears: grains that have an
even number of electrons, $N$, have the lower energy than those
that have an ``unpaired" odd electron that carries extra energy
$\Delta$ \cite{comment1}.  We demonstrate that these features
affect strongly transport properties of granular superconductors
when electrostatic disorder is taken into account. Depending on
the relation between the local charging energy, $E_c$, and the
local superconducting gap, $\Delta$, the low temperature transport
goes either via hopping of electrons for $E_c>\Delta$ (we
hereafter will be referring to this mechanism as to EH regime) or
via hopping of the Cooper pairs for $E_c<\Delta$ (CPH regime). We
show that this classification holds as long as the dispersion in
grain sizes can be neglected; neither it is affected by the
presence of the of-diagonal part of the Coulomb interaction.
Derivation of the conductivity temperature behavior in a general
case of an arbitrary relation between $E_c$ and $\Delta$ is,
however, cumbersome and below we present the results only in the
limiting cases corresponding to $E_c\gg \Delta$ in EH- and to
$E_c\ll\Delta$ in CPH regimes.

In the EH regime the conductivity,  is determined either by the
elastic or inelastic cotunneling mechanism depending on the
temperature regime. In the low temperature elastic regime, the
conductivity is almost unaffected by the gap at $E_c \gg \Delta$
and follows the ES law \cite{Zhang05,BLV05,Feigelman05}:
\begin{equation}
\label{ES} \sigma \sim \exp[-(T_0/T)^{1/2}], \;\;\;\;
 T_0 = b\, e^2/ \tilde
\kappa\, a\, \xi^{EH},
\end{equation}
where $e$ is the electron charge, $\tilde\kappa$ is the effective
dielectric constant of a granular array
\cite{dielectric_constant}, $\xi^{EH}$ is the localization length
measured in the units of the grain diameter $a$ (we consider a
model where all grains are of the same size) and $b$ is the
numerical factor of the order of unity depending on the array
morphology. The localization length in the case of elastic
cotunneling is \cite{BLV05}
\begin{equation}
\label{localization} \xi_{el}^{EH} =  2/ \ln (\, \bar  E \, \pi /
\bar g \, \delta ),
\end{equation}
where $\bar g$ is of the order of the average tunneling
conductance; more precisely it is the geometrical average of the
tunneling conductances $g_{i,i+1}$ along the typical tunneling
path $\ln \bar g = \langle \ln g \rangle \equiv
(1/N)\sum_{i=1}^N{\ln g_{i,i+1}}.$ The energy $\bar E \sim E_c$ is
analogously defined as a geometrical average $\ln \bar E = \langle
\ln \tilde E \rangle,$ where $\tilde E$ is the combination of
single grain electron $ E_+^i$ and hole $ E_-^i$ excitation
energies at $\Delta=0$
\begin{equation}
\tilde  E^i = 2/(1/ E_+^i + 1/E_-^i).
\end{equation}

In the absence of the superconducting gap the elastic transport
regime crosses over at $T \approx T_1$ to the regime dominated by
inelastic cotunneling processes  where the conductivity is given
by ES law with weakly temperature dependent localization length
\cite{BLV05,Feigelman05}
\begin{equation}
\label{inelastic_loc_len} \xi^{EH}_{in} = 2 / \ln ( \bar E^2/ 16
\pi T^2 \bar g ).
\end{equation}

In the presence of the gap and at $T\ll\Delta$ the inelastic
cotunneling processes are suppressed.  Thus the result (\ref{ES})
holds only in the interval  $T\gtrsim {\rm max}\, \{ \Delta, T_1
\}$.  Considering the case $T_c > T_1$, we find the new transport
regime at $T<T_3\approx \xi_{in}^{EH}\Delta$, which is dominated
by inelastic processes but has the activation form:
\begin{equation}
\label{crossover} \sigma \sim \exp\Big[-N \big(\ln(\bar E^2/ 4
{\bar g}  T\Delta ) +2\Delta/T \big) \Big],
\end{equation}
where the typical tunneling order $N=\sqrt{ b \, e^2/ 16\,
a\tilde\kappa\Delta} \sim \sqrt{E_c/\Delta}.$ Under temperature
decease dependence (\ref{crossover}) remains applicable till it
matches the elastic result (\ref{ES},\ref{localization}) at
temperature $T_2 \approx  \xi^{EH}_{el} \Delta.$

Changing the gap $\Delta$ by varying magnetic field at fixed
temperature one can drive the system between the elastic and
inelastic ES regimes traversing the interval $T_2(H)<T<T_3(H)$,
where conductivity follows the activation formula of
Eq.(\ref{crossover}) (see Fig.1).  This manifests itself as a
giant negative magnetoresistance in the temperature interval
$(T_1, \,T_3)$ that exists provided $T_1 \ll T_c$. Such an effect
in the weak coupling regime was indeed observed in
Ref.\cite{Gerber}.

In the CPH regime the  conductivity follows the ES law with $T_0$
and the localization length given by
\begin{equation}
\label{xi_B} T_0 = b\, (2e)^2/ \tilde \kappa\, a\, \xi^{CPH},
\hspace{0.4cm}
 \xi^{CPH} = 1/\ln(8 \bar E / \pi \bar g \Delta).
\end{equation}
We see that the localization length decreases with decrease of
$\Delta$, which corresponds to the strong positive
magnetoresistance.   Note further that at $\bar g \Delta \sim E_c$
the sample transforms into a superconducting state\cite{Efetov80}.

We begin our quantitative analysis with the discussion of the
ground state charge configuration at $T\to 0$. First we consider
an {\it isolated} superconducting grain in the presence of the
random potential $V$ modelling electrostatic disorder. The energy
of such a grain is
\begin{equation}
\label{model1} E = n^2 E_c - V n +   P(n+p)\,\Delta ,
\end{equation}
where $P$ is the parity function defined as $P(2k+1)=1,\, P(2k)=0$
for an integer $k$, and $n$ is the number of excessive electrons,
counted with respect to the number of electrons, $N^0$, of a
neutral state at $V=0$, $\Delta=0$. The discrete random variable
$p=0$ for even parity and $p=1$ for odd parity of the charge
neutral state $N^0$.

The effect of the applied potential $V$ on the ground state is
qualitatively different for different relations between $E_c$ and
$\Delta$~\cite{Averin_Nazarov2,Matveev}. If $E_c>\Delta$ (EH), the
dependence of the ground state charge on $V$ has a form of the
Coulomb staircase with the charge jumps $n \to n+1$ at $V_n=
(2n+1)E_c + \Delta \cos\pi(n+p)$. In the case $E_c<\Delta$ (CPH),
the dependence of the occupation number on $V$ also has a
staircase form but electrons jump in pairs, $n\to n+2$, at the
equidistant values $ V_n= E_c(2n+2)$, with even-, $n=2k$, for
$p=0$ - and odd, $n=2k+1$, for $p=1$, occupation numbers ($k$ is
an integer).

\begin{figure}[tbp]
\vspace{-0.3cm}
\includegraphics[width=2.8in]{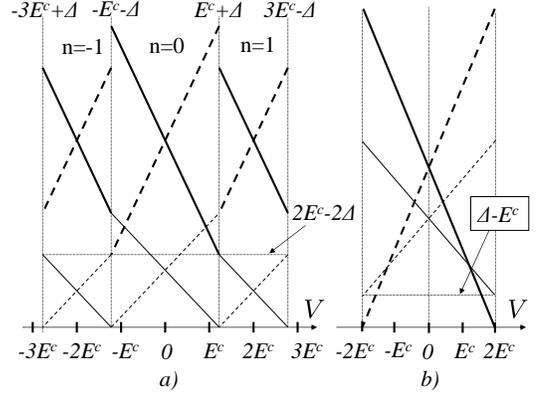}
\vspace{-0.4cm}
 \caption{
 Single particle (thin lines) and two particle (thick
lines) excitation energies as functions of the applied potential
in a single grain for the cases a) $E_c>\Delta$ and b)
$E_c<\Delta$. Solid lines represent creation processes (addition
of an extra electron or a pair) while dashed line represent
annihilation processes. In both cases the dependencies are $4E_c$
periodic.
\vspace{-0.5cm}
 }
  \label{Spectrum1}
\end{figure}

Further, one can see that in the EH regime the Cooper pair
excitations have a gap, while single particle excitations are
gapless for certain values of $V$. On the contrary, in the CPH
regime electron and hole excitations are gapped for all potential
values while the pair excitations are gapless at certain values of
$V$. Shown in Fig.1 are electron ${\cal E}_+$ and hole ${\cal
E}_-$ excitation energies
\begin{equation}
{\cal E}_{\pm} =(\pm 2n+1)E_c \mp V+\Delta \cos\pi (n+p),
\end{equation}
and pair creation ${\cal E}_{2+}$ and annihilation ${\cal E}_{2-}$
energies
\begin{equation}
{\cal E}_{2\pm} = 4(\pm n+1)E_c \mp 2V,
\end{equation}
as functions of $V;$ the minimal pair excitation energy in the EH
regime is $2(E_c-\Delta)>0,$  and the minimal energy of an
electron excitation in the CPH regime is $\Delta-E_c>0$.

We assume that electrostatic disorder is strong taking
characteristic dispersion of $V$ as $V_0 \gtrsim E_c$. Averaging
over the potential $V$ results in the finite density of low energy
electron-hole excitations in the EH regime while in the CPH regime
only the pair gapless excitations appear.

The excitation spectrum we have discussed refereed to
noninteracting granules.  Now we introduce interactions:
\begin{equation}
\label{real_model} H^c = \sum_{ij} n_i E_c^{ij} n_j - V_i n_i+
P(n_i+p_i)\, \Delta,
\end{equation}
where the matrix $E_c^{ij}$ represents the Coulomb interaction
related to the capacitance matrix $C_{ij}$ as $E_c^{ij}=(e^2/2)
C_{ij}^{-1}.$  Remarkably, the main conclusion concerning the
nature of low energy excitations depending on the ratio of $E_c
\equiv E_c^{ii}$ and $\Delta$ obtained for a single grain holds
for model (\ref{real_model}). Indeed, the energies of the single
and two particle excitations on the site $i$ following from
(\ref{real_model}) are
\begin{eqnarray}
&{\cal E}_{\pm}^i &= \pm \mu_i + E_c + \Delta \cos\pi (n_i+p_i),  \label{first}  \\
&{\cal E}_{2\pm}^i &=  \pm 2\mu_i + 4 E_c,  \label{second}
\end{eqnarray}
where the local potential $\mu_i$ is $\mu_i = 2 \sum_j E_c^{ij}
n_j - V_i.$  The stability of the ground state charge
configuration assumes that ${\cal E}_{\pm}^i, {\cal E}_{2\pm}^i >0
$ that limits possible values of $\mu_i$. Now suppose that $\mu_i$
is such that an electron excitation on the grain $i$ is gapless: $
{\cal E}_+^i=0.$  The ground state stability requires that $ {\cal
E}_-^i,\, {\cal E}_{2+}^i,\, {\cal E}_{2-}^i
> 0 $ giving immediately  $E_c>\Delta$ and ${\cal E}_{2\pm}^{i}
> 2(E_c-\Delta).$ This means that gapless electron
excitations exist only in the case $E_c > \Delta$ and that at the
same time pair excitations have a minimal gap $2(E_c-\Delta).$
Analysis of the case ${\cal E}_{-}^i=0$ yields the same result.

Similarly, for a gapless pair excitation to appear at a certain
grain $i$, it is required that $ {\cal E}_{2+}^i=0$ and stability
assumes ${\cal E}_-^i,\, {\cal E}_{+}^i,\, {\cal E}_{2-}^i > 0,$
meaning that $n_i+p_i$ has to be even and that $\Delta> E_c,\;
{\cal E}_{\pm} > \Delta -E_c.$ This in tern assumes that the
gapless pair excitations exist only in the case $E_c<\Delta$ and
that single particle excitations has the gap $\Delta-E_c
> 0$.

\begin{figure}[tbp]
\includegraphics[width=3.2in]{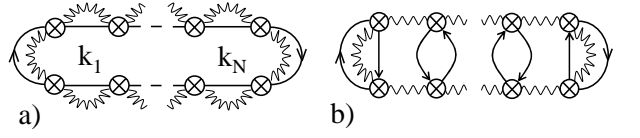}
\caption{ The diagrams represent the tunneling probability via
elastic (a) and inelastic (b) cotunneling processes. The crossed
circles stay for the tunneling matrix elements
$t^{ij}_{k,k^\prime}\, e^{i\phi_i(\tau)} \, e^{-i\phi_j(\tau)}$
where phase factors appear due to the gauge transformation. Wavy
lines represent the average of the phase factors $\langle
e^{i\phi(\tau_1)} \, e^{-i\phi(\tau_2)} \rangle$ with respect to
the Coulomb action. }
 \vspace{-0.3cm}
  \label{Diagram}
\end{figure}

 To find the density of low energy excitations we follow the
 standard ES approach~\cite{ES}. For the EH regime we
 require that the energy of the excitation consisting of replacing an electron
 from the site $i$ to the site $j$ were positive:
 $
 {\cal E}_{-+}^{ij} = {\cal E}_-^i+{\cal E}_+^j - 2E_c^{ij} >0,
 $
 resulting in the standard ES expression for the low energy single particle density of
 states (DOS)
\begin{equation}
\label{nu_1}
\nu_1(\varepsilon) = \alpha_{d1}\, ( \tilde\kappa/e^2
)^d \varepsilon^{d-1},
\end{equation}
where $\alpha_d$ is the dimensionless factor \cite{comment2}. In
the CPH regime we find the density of two particles excitations by
requiring the stability of the ground state with respect to a
replacement of a pair from site $i$ to the site $j$: $ {\cal
E}_{2-2+}^{\;i\;\;j} = {\cal E}_{2-}^i+{\cal
E}_{2+}^j-8E^{ij}_c>0. $ We again arrive at the ES stability
condition leading to the pair DOS as
\begin{equation}
\nu_2(\varepsilon) = \alpha_{d2}\, ( \tilde\kappa/(2e)^2 )^d
\varepsilon^{d-1}.
\end{equation}
The two regimes differ only in the doubled charge of the Cooper
pair, and thus we expect that $\alpha_{d1}\approx \alpha_{d2}$.

Now we turn to the derivation of hopping probabilities. We work in
the basis of exact eigenstates of noninteracting isolated grains.
Electron tunneling processes between the states $k_i$ of the grain
$i$ and $k_j$ of the grain $j$ are represented by the elements of
the tunneling matrix $t^{ij}_{k_i,k_j}$.  The Coulomb interaction
within each grain is accounted for via the gauge transformation of
the electron fields $c_k(\tau)\to c_k(\tau) \,e^{i\phi(\tau)},$
and the phase field $\phi$ appears as a renormalization of the
hopping matrix elements \cite{note3}
\begin{equation}
t_{ij} \to t_{ij} \, e^{i\phi_i(\tau) -i\phi_j(\tau) }.
\end{equation}
The phase field $\phi$ is governed by the Coulomb action
\begin{equation}
S=-{1\over {2 e^2}}\sum_{ij} \int d\tau \, (\dot \phi_i +i V_i) \,
C_{ij} \, ( \dot \phi_j +i V_j).
\end{equation}
Without the Coulomb interaction a grain is described by the
standard Bardeen-Cooper-Schrieffer model.

The tunneling probability via the elastic process  given by the
diagram a) in Fig.~(\ref{Diagram}) can be written as a product $ P
\sim \prod_i g_{i,i+1}\, P_i  $ with $P_i$ representing an
elementary contribution from the grain $i$
\begin{equation}
\label{P_i}
 P_i = {{\delta}\over {2\pi}}  \int\,d\xi d\tau_1 d\tau_2 G_\xi(\tau_1)\,
G_\xi(\tau_2)  e^{ -E^i_+|\tau_1|-E^i_-|\tau_2| }.
 \end{equation}
Here $E_+^i$ and $E_-^i$ are given by Eq.(\ref{first}) with
$\Delta=0$ and $G_\xi(\tau)$ is the superconducting Green function
\begin{eqnarray}
G_\xi(\tau)=\mp{1\over 2} \Big[ f(\mp E_\xi/T)\, (1+\xi/E_\xi )\,
e^{-E_\xi \tau}
 \nonumber \\
 + f(\pm E_\xi/T)\, (1-\xi/E_\xi )\, e^{E_\xi\tau}  \Big],
\end{eqnarray}
where the upper(lower) sign stays for $\tau>0$ ($\tau<0$),
$f(x)=1/(1+e^x)$ and $E_\xi=\sqrt{\xi^2+\Delta^2}.$ Taking the
integrals in (\ref{P_i}) in the limit $T\ll\Delta$ we obtain
\begin{equation}
P_i = \delta /\pi \tilde {\cal E}^i, \;\;\; \tilde {\cal E}^i =
\tilde E^i +\pi \Delta/2,
\end{equation}
that leads to the localization length of the dependence
(\ref{ES}). Neglecting the term $\pi\Delta/4$ in the expression
for $\tilde {\cal E }^i$ we arrive at the expression
(\ref{localization}) for the localization length. We would like to
note that this correction though being small in the considered
limit ($E_c\gg \Delta$) can still lead to a noticeable negative
contribution to the magnetoresistance in the regime of elastic
cotunneling.

The tunneling probability via inelastic cotunneling processes is
given by the diagram b) in Fig.~(\ref{Diagram}) that in the limit
$T,\Delta \ll E_c$ gives
\begin{equation}
\label{P_{in}} P_{in} \sim (4\bar g/\pi \bar E^2 )^N
\int_{-\infty}^\infty\, dt\, [\,G(-it)\, G(it)\, ]^{N} \, e^{-i
\varepsilon t },
\end{equation}
where $N$ is the tunneling order, $G(\tau)= \int \, d\xi\,
G_\xi(\tau)$ and $\varepsilon$ is the electron energy change. The
expression (\ref{P_{in}}) has to be minimized with respect to $N$
under constraint $N a \varepsilon \tilde \kappa/e^2\sim 1,$ that
follows from the Mott argument for finding the typical minimal
distance to a site available for electron placement within the
energy shell $\varepsilon$ and from ES expression (\ref{nu_1}) for
DOS. In the limit $N \gg 1$ the integral in (\ref{P_{in}}) can be
taken within the saddle point approximation leading to ES law
(\ref{ES},\ref{inelastic_loc_len}) for temperature $T > T_3$ and
to the activation behavior (\ref{crossover}) at  $T < T_3.$

Deriving the result (\ref{crossover}) we neglected the possibility
of inelastic cotunneling through the unpaired states
\cite{Averin_Nazarov2} of "odd" grains that constitute about a
half of all grains in the case $E_c\gg \Delta.$ One can show that
such processes do not affect the results since the elementary
probability of the inelastic cotunneling process via an unpaired
state is smaller that for the elastic process by the factor of
$T/E_c.$

The tunneling in CPH regime can be described within the effective
model acting on Cooper pairs
 \begin{equation}
 \nonumber
 H = 4 \sum_{ij} \hat n_i E_c^{ij} \hat n_j - 2 \sum_i \hat n_i V_i +
 (1/2) \sum_{<ij>} J_{ij}\, e^{i\varphi_i-i\varphi_j} ,
 \end{equation}
where $\varphi$ and $ \hat n = -i\partial/\partial \varphi $ are
the Cooper pair phase and number operator respectively and
$J_{ij}=g_{ij}\pi\Delta/2$ \cite{AB} is the Josephson coupling
between the neighboring grains $i$ and $j.$ Tunneling amplitude,
$A$, can be calculated within the straightforward perturbation
theory in $J$:
\begin{equation}
A \sim \prod_{i=1}^N \, J_{i,i+1}/\tilde {\cal E}^i_2,  \;\;\;
\tilde {\cal E}_2^i= 2/[\,{1/\cal E}_{2+}^i +1/{\cal E}_{2-}^i],
\end{equation}
where pair excitation energies ${\cal E}_{2+}^i, {\cal E}_{2-}^i$
are defined by Eq.~(\ref{second}). The tunneling probability,
$P=A^*A$,  decays with the tunneling order $N$ as
$e^{-2N/\xi^{CPH}}$ with the localization length $\xi^{CPH}=1/\ln(
2 \bar E_2 / \pi \bar g \Delta)$, where $\bar E_2$ is the
geometrical average of effective pair excitation energies $\tilde
{\cal E}_2^i$ along a typical tunneling path. Since the density of
states in EH and CPH regimes differ only by the effective charge,
$\bar E_2\approx 4 \bar E$ leading to Eq.(\ref{xi_B}).

In conclusion, we have described the hopping transport in granular
superconductors and found that if the single grain charging energy
exceeds the superconducting gap, $E_c >\Delta,$ the transport goes
via hopping of electrons, while in the opposite case, $E_c <
\Delta$, hopping of Cooper pairs dominates the transport. In the
former case we predict the negative magnetoresistance, while the
latter regime exhibits the positive magnetoresistance. We relate
the giant negative magnetoresistance observed in the insulating
phase of the granular superconductors in \cite{Gerber} to the
suppression of the inelastic cotunneling in the electron hopping
dominated regime. Transport via Cooper pair hopping can be
observed in samples with low enough grain charging energy, i.e.
large enough grains. Such regime can also appear as a result of
the renormalization of the Coulomb energy due to intergranular
coupling in samples with intermediate coupling strength $g \sim
1,$ in particular close to the insulator to superconductor
transition one expects CP dominated transport. The regime $g\sim
1$ is a subject of our future investigation.

We would like to thank I.~S.~Beloborodov, Yu.~M.~Galperin and
H.~M.~Jaeger for useful discussions. This work was supported by
the U.S. Department of Energy, Office of Science via the contract
No. W-31-109-ENG-38.


\vspace{-0.4cm}

\begin{thebibliography}{99}

\bibitem{Gerber} A.~Gerber \textit{et al, } Phys. Rev. Lett.~\textbf{78}, 4277
(1997), and references therein.

\bibitem{Abeles} P.~Sheng, B.~Abeles, and Y.~Arie, Phys. Rev. Lett.~\textbf{31}, 44
(1973); B.~Abeles, P.~Sheng, M.~D.~Coutts, and Y.~Arie, Adv.
Phys.~\textbf{24}, 407 (1975).

\bibitem{Zhang05} J. Zhang and B. I. Shklovskii, Phys. Rev. B {\bf 70}, 115317
(2004).

\bibitem{BLV05} I.~S.~Beloborodov, A.~V.~Lopatin, and
V.~M.~Vinokur, Phys. Rev.  B {\bf 72} 125121 (2005).

\bibitem{Feigelman05} M.~V.~Feigel'man and A.~S.~Ioselevich,
Pis'ma Zh. Eksp. Teor. Fiz. {\bf 81}, 341 (2005) [Sov. Phys. JETP
Lett. {\bf 81}, 227 (2005)].

\bibitem{ES} B.~I.~Shklovskii and A.~L.~Efros, {\it Electronic
properties of Doped Semiconductors, Springer-Verlag, New York,
1988)}.

\bibitem{Averin} D.~A.~Averin and Yu.~V.~Nazarov, Phys. Rev. Lett.
\textbf{\ 65}, 2446 (1990).

\bibitem{Jaeger}  T.B.Tran et. al., Phys. Rev. Lett. {\bf 95}, 076806 (2005).

\bibitem{book} Eugen ${\rm\breve{S}}$im${\rm \acute{a}}$nek, {\it Inhomogeneous
Superconductors}, Oxford University Press, 1994.

\bibitem{Averin_Nazarov2}  D. V. Averin and Yu. V. Nazarov,  Phys. Rev. Lett. {\bf 69},
1993(1992).

\bibitem{Matveev} K. A. Matveev et. al., Phys. Rev. Lett. {\bf
70},
2940 (1993).

\bibitem{comment1} We assume that $ \Delta \gg \delta$ and neglect
 the energy contribtions to a single grain of the order of
 $\delta.$

\bibitem{dielectric_constant}  Note that $\tilde\kappa$ can be
much larger than the dielectric constant of the insulating
component of a sample, see Ref.\cite{Zhang05}.

\bibitem{Efetov80} K.~B.~Efetov, Sov. Phys. JETP~\textbf{51}, 1015
(1980), E. ${\rm\breve{S}}$im${\rm \acute{a}}$nek, Phys. Rev. B
{\bf 22}, 459 (1980).

\bibitem{comment2} Strictly speaking $\nu_1(\varepsilon)$
represents the density of {\it ground} states, for discussion see,
for eaxmple, Refs.\cite{Zhang05,BLV05}.

\bibitem{note3} For detials see for eaxample Ref.~\cite{BLV05}.

\bibitem{AB} V. Ambegaokar and A. Baratoff, Phys. Rev. Lett. {\bf
10}, 486 (1963).




\end{thebibliography}
\end{document}